# Hybrid Nano Communications

S. M. Riazul Islam, M. Jalil Piran, and Anish Prashad Shrestha
Dept. of Computer Science and Engineering
Sejong University
Seoul, South Korea
{riaz, piran, anishpshrestha}@sejong.ac.kr

Farman Ali, Asif Iqbal, and Kyung-Sup Kwak
UWB Wireless Communications Research Center
Inha University
Incheon, South Korea
{farmankanju, asifsoul}@gmail.com and kskwak@inha.ac.kr

*Abstract*—In this paper, we propose a hybrid nano communication (HNC) system that integrates three possible modes of nano communications: terahertz communication (TC), molecular communication (MC), and neural communication (NC). The paper provides some insights into how the required four key building blocks, namely, terahertz to molecular (T2M) relay interface, molecular to terahertz (M2T) interface, molecular to neural (M2N) relay, and neural to molecular (N2M) interface can be designed to implement the HNC system. In addition, the capacity analysis for the hybrid channel is provided. Sub-channels' capacities are also calculated. Finally, numerical results are presented to comprehend the data transmission capacity of the proposed HNC system.

## I. Introduction

In many applications where electromagnetic (EM) communication is not convenient, molecular communication (MC) is being considered as an alternative approach [1]. This emerging communication paradigm uses molecules as information carriers. To put it simply, a transmitter in a MC system releases tiny particles (e.g., molecules or ions) into a fluidic or gaseous medium by means of diffusion or drift mechanism. The particles propagate and eventually reach to a receiver. After detecting the particle, the receiver then decodes the information bits encoded in the transmitted particles. Some of the representative applications of MC are networks of tunnels and pipelines, underwater environments, nano-scale medical robots, and environmental monitoring. Another immense area of MC applications is healthcare. In fact, healthcare is about to experience a paradigm shift from traditional physician-centric models to patient-centric approaches. The integration of various new technologies such as wearables, sensors, and internet of things (IoT) is propelling the notion of patient empowerment offering a new form of disease prevention, diagnose and treatment. One of the fundamental interests of today's healthcare stakeholders is to develop various healthcare solutions which would be minimally invasive and biocompatible. MC is thus highly potential to be adopted as one of the core technologies for implementing such advanced medical solution [2].

Despite its potentials, the MC mode of nano communication alone is not suitable to implement an effective communication network for some contexts such as human body, a complex and huge environment. Because of inherent limitation on motion of diffusion-based communication, MC mode cannot be used as an effective mean for body area network (BAN) to BAN, inter-BAN, communications. Compared to the MC mode, terahertz (THz) mode of nano communications [3] comes with better coverage and directivity making an appealing mean for device-to-device (D2D) communications in general and for inter-BAN and intra-BAN in particular. Keeping the THz communications (TC) apart, from physiology perspective, diffusion-based communication is slower compared to neural communication (NC) in a neural network (NN) [4]– the electrical/chemical signal transmission from one neuron to another occurs faster via a synapse. Moreover, although a nano-machine (NM) in a MC system comes with a higher mobility than a synapse in the nano communication system, the NM cannot move everywhere. Rather, to take the benefits of both NM and synapse, it would be more sensible to integrate them in a single system. Taking together, the three modes of nano communications – TC, MC, and NC – pointed above can be combined together using some appropriate interfaces so that inter-BAN, intra-BAN, in-body nano communications, and neural network can be viewed as a single system. We term that single system Hybrid Nano Communication (HNC). In this paper, we elaborate the different elements such as required interfaces and channels associated with the proposed HNC system.

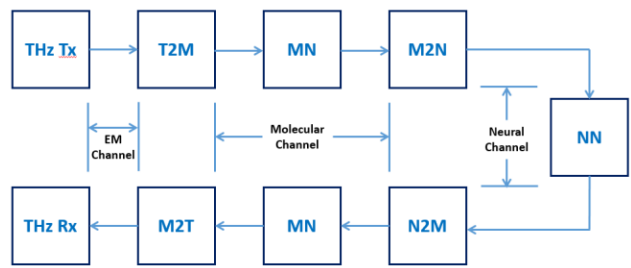

**Fig. 1** Hybrid nano communication (HNC) system model (T2M: terahertz to molecular, MN: molecular network, M2N: molecular to neural, NN: neural network, N2M: neural to molecular, M2T: molecular to terahertz).

This work was supported by National Research Foundation of Korea-Grant funded by the Korean Government (Ministry of Science and ICT)-NRF-2017R1A2B2012337.

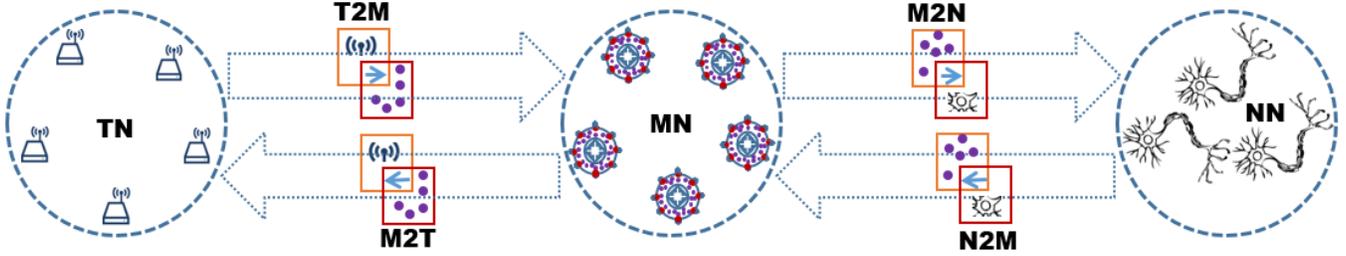

**Fig. 2** Functional diagram of the HNC System.

## II. HNC System

As presented in Fig. 1, the HNC system primarily comprises four different types of interfaces and three different kinds of channels. The terahertz to molecular (T2M) and molecular to terahertz (M2T) interfaces connect EM channel with molecular channel. Whereas the former converts electromagnetic signal to molecular bit streams, the latter does the opposite i.e., it transforms molecular information to the terahertz signal. Of the remaining interfaces, the molecular to neural (M2N) interface is responsible to regulate molecular concentration which eventually modulates the release of neurotransmitters, whereas the neural to molecular (N2M) interface decodes the neural signal and convert it to molecular information bit streams. As such, M2N and N2M interfaces connect molecular channel with neural channel. The HNC system can be interpreted as an integrated framework combining THz network (TN) where nano devices communicate with each other over EM channel, molecular network where NMs communicate with each other via molecular channel, and neural network where neurons communicate via synapses.

A functional diagram of the HNC system is provided in Fig. 2. The TN can be an on-body or out-body network surrounding a human body like wireless body area network (WBAN). Even, a TN can be interconnected with another TN forming inter-WBAN. As such, the location of a THz nano device is usually outside of a human body. An MN is formed by a group of neighboring NMs with or without mobility. Multiple NMs in a MN covering a particular area inside the human body are used to precisely access the NN in the vicinity. To cover the entire human body, multiple MNs are therefore required, each of which targets the respective local part of the NN. Suppose that, a THz device in a TN aims to communicate with a molecular NM located at a distant MN inside the human body. Suppose that, a THz device in a TN aims to communicate with a molecular NM located at a distant place inside the human body. The THz device transmits its intended information bits using EM signal. Based on the available routing information, the assigned T2M relay forwards the EM signal to the expected MN by converting the signal into molecular form. The molecular bits stream eventually reaches to the assigned M2N relay that activates neurotransmitters of synapse in a controlled manner by regulating the concentration of molecules. Next, the neural signal passes a longer distance through the connected synapses over the NN and ultimately reaches to the expected N2M relay that in turn decodes the neural signal and forwards the signal to the expected MN in which the distant NM is connected with. It is worth mentioning that although it is possible to forward the molecular signal from the distant MN to another out/on-body TN by converting the molecular information bit streams to EM signal through M2T interface although the case is less likely because out/on-body communications does not necessarily require MN and NN.

## III. HNC Sub-Syetms

Based on the system overview provided in the preceding section, it is obvious that the four interfaces, namely, T2M, M2T, M2N, and N2M are the most important building blocks for implementing the proposed HNC system. In this section, we provide the details of how these subsystems can be realized.

*A. T2M Relay:*

The diagram presented in Fig. 3 shows a simple architecture of the T2M interface. The capacitor C2 is charged when a nano pulse is captured by nano antenna. C2 charges up to the peak value of the nano pulse received through the diode D1. The capacitor C1 is used to block any direct current. When C2 reaches a threshold voltage, it turns on the UJT U1 which in turn triggers the power MOSFET M1 so that required current flows through a nanofluidic device. The nanopump pushes molecule reservoir to release molecules through appropriate channel. The detailed design of the nanopump is beyond the scope of this paper; the interested reader is referred to [5] for some insights into how electroosmotic flow can be utilized as a pumping technology. It can be noted that once UJT is switched on C2 discharges though R2 so that it can again be charged by next nano pulse. Band-pass filter (BPF) can be tuned to specific frequency band.

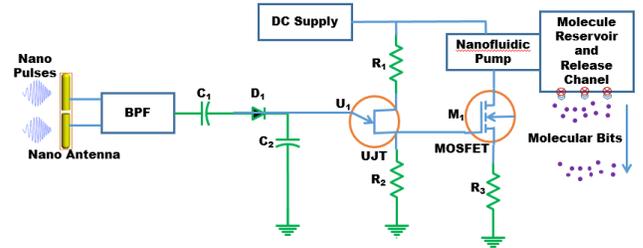

**Fig. 3** T2M relay converts THz wave to molecular bit streams (the value of each circuit element depends upon a particular design aspect)

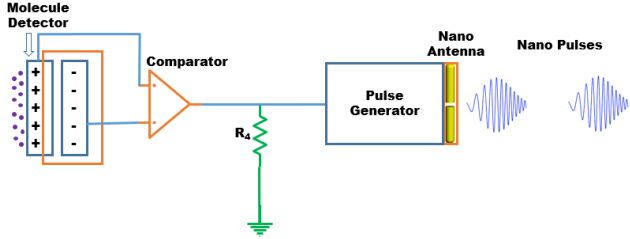

**Fig. 4** M2T relay converts molecular information to THz wave.

*B. M2T Relay:*

The working principle of a M2T relay can be described with the help of Fig. 4. The molecule detector is a cathode-anode-like arrangement. When a sufficient number of messenger molecules reaches molecule detector, a threshold potential difference builds up across the detector. This voltage is sensed by the comparator and its output become 1. The comparator output basically acts as a switch of a pulse generator. When, the output of the comparator becomes 1, the generator is switched on and a Gaussian nano pulse is transmitted. As far as the pulse generator is concerned, all of the components except a required radio frequency (RF) inductor can be implemented on a single complementary metal–oxide–semiconductor (CMOS) chip [6], while the RF choke needs to be off-chip.

*C. M2N Relay:*

The schematic diagram shown in Fig. 5 depicts how the M2N relay works. The M2N has two parts. The first part consists of a molecule detector and a comparator, whereas the other part consists of the UJT, MOSFET, nano-pump and ion-reservoir and channel arrangement. The first part of the M2N is alike to the first part of M2T relay while the other part is analogous to the second part of M2T relay. On that, the operation of M2N can be summarized by looking into the sensing input of M2T followed by the response of T2M. As such, the comparator's output eventually cause the releasing of $ca^{2+}$ from the ion reservoir. The variation in ion concentration modulates the chemical and electrical properties of synapse, explained in the next subsection.

*D. N2M Relay:*

This interface (see Fig. 6) is the most critical part of the HNC system. Although the complete mechanism of N2M is yet to be elucidated, we try to provide some insights into a possible working principle. Calcium influx, originated by a nearby M2N relay, into the presynaptic end controls the release of vesicles. In fact, the variation in Ca2+ concentration is detected by presynaptic neuron using appropriate sensors such as synaptotagmin. The sensors thus allow synaptic vesicles to release neurotransmitter molecules in a controlled manner. There exists a number of synaptic vesicles, each of which comes with a certain release probability due to Ca2+ influx. It can be noted that each vesicle contains a certain amount of molecules as neurotransmitters [7]. Neurotransmitter molecules eventually reach and bind to ionotropic receptors through diffusion acting as ion channels that directly regulate the electrical properties of postsynaptic neuron. The altered electrical properties need to be tr anslated into information bit streams. To do so, the first step is to transform the postsynaptic waveform to axonal impulse train and therefore the output of postsynaptic neuron would be obtained by taking the convolution of the impulse train and response function [8]. Next, the binary sequence is extracted from the impulse train and the binary information is then obtained by testing the sequence distribution. The binary output triggers the expected molecular nano machine which releases molecular bit streams.

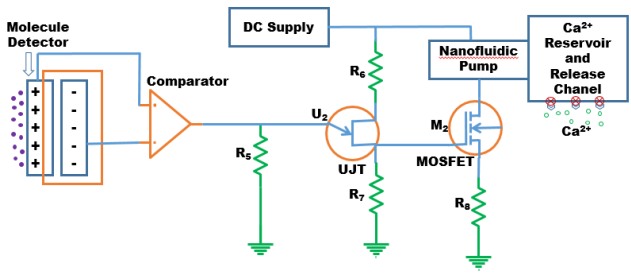

**Fig. 5** M2N relay converts molecular information to the varying ion concentration pattern stimulating neurotransmitters.

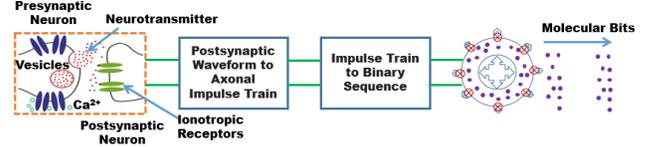

**Fig. 6** N2M relay decodes $Ca^{2+}$-induced neural signal variation to molecular information bit streams.

## IV. CAPACITY ANALYSIS

Fig. 7 depicts data transmission using hybrid channel. The hybrid channel model however does not include neural to molecular transmission based on the assumption of channel reciprocity where molecular to neural channel would be identical to neural to molecular channel. Whereas the THz end of the T2M relay has been absorbed into the EM channel, the molecular end of channel has been considered together with the molecular channel. Consequently, the output variable of EM channel $S'$ remains the same as the input variable of molecular channel. In the same way, the variable $S''$ is considered both the output of molecular channel and input of the neural channel as there is ideally no loss of information in M2N interface. If the mutual information between the random variables $X$ and $Y$ be $I(X,Y)$, the channel capacity of the hybrid channel can be expressed as

$$C_H = \max_X \{I(X,Y)\} \quad (1)$$

Equation (1) can be tailored to calculate the capacities of sub-channels. The capacity of the EM sub-channel $C_1$, which is the

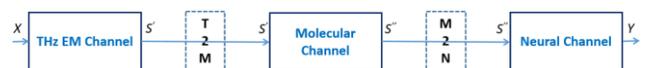

**Fig. 7** Hybrid channel consisting of EM channel, molecular channel and neural channel.

maximum mutual information between random variables $X$ and $S'$, can be written as

$$C_1 = \max_X \{I(X, S')\} \quad (2)$$

THz channel is highly frequency selective. Therefore, the system bandwidth need to be divided into a number of sub-bands [9] and the channel capacity can thus be calculated by summing the individual capacities as

$$C_1 = \sum_i \Delta f \log_2 \left[1 + \frac{S(f_i) A(f_i, d_1)^{-1}}{N(f_i, d_1)}\right] \quad (3)$$

$f_i$ is the center frequency of the THz sub-band $i$, $d_1$ is distance between a THz transmitter and a T2M relay as the respective receiver, $S$ is transmitted signal power spectral density (PSD), $A$ is the path loss, and $N$ is noise PSD.

If we analyze and manipulate the mutual information $I(S', S'')$ following the steps carried out in [10], the capacity for diffusion-based molecular channel can be written as

$$C_2 = 2W\left(1 + \log_2 \frac{\bar{P}}{3WK_BT}\right) - 2\log_2(\pi D d_2) - \frac{4d_2}{3\ln 2}\sqrt{\frac{\pi W}{D}}$$
$$+ 2W \frac{2\bar{P}R_d}{9W^2 d_2 K_B T} - 2W\ln(W\tau) - 2W\ln\left(\Gamma\left(\frac{2\bar{P}R_d}{9W^2 d_2 K_B T}\right)\right)$$
$$- 2W\left(1 - \frac{2\bar{P}R_d}{9W^2 d_2 K_B T}\right)\psi\left(\frac{2\bar{P}R_d}{9W^2 d_2 K_B T}\right) \quad (4)$$

$W$ is the bandwidth of the signal $S'$, $\bar{P}$ is the mean transmit thermodynamics power, $K_B$ is the Boltzmann constant, $T$ is the absolute system temperature, $D$ is diffusion coefficient, $d_2$ is distance between a molecular NM transmitter and a M2N relay as the respective receiver, $R_d$ is the radius of molecular detector, $\tau$ is the time interval associated with molecule distribution, and $\psi(\cdot)$ is the digamma function.

For a neuron refractory period, minimum time required for a neuron to be responsive between two consecutive stimuli, of $\delta$, the channel capacity of a single neuron is calculated in [11]. Therefore, if a neural link consists of a series of similar neurons, the channel capacity for the link with Gaussian noise can be written as

$$C_3 = \frac{aH}{1 + a\delta} \quad (5)$$

where $a$ is the average input rate in pulse per second and $H$ is the information per signal that can be expressed as

$$H = a\sigma e^{-a\sigma} - (1 - e^{-a\sigma})\ln(e^{-a\sigma}) \quad (6)$$

where $\sigma$ is the standard deviation of latency. The capacity of the cascaded connection thus becomes

$$C = min\{C_1, C_2, C_3\} \quad (7)$$

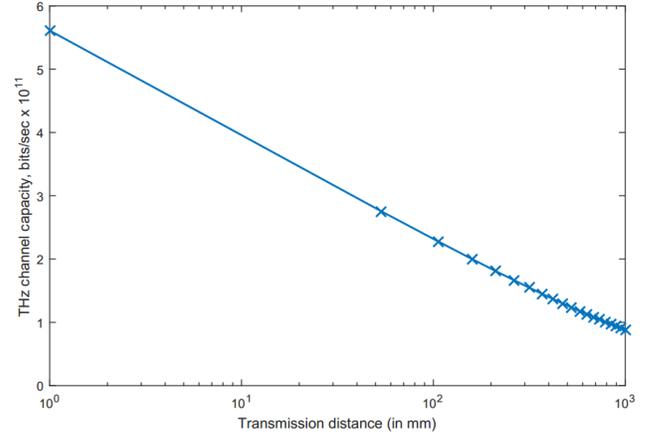

**Fig. 8** Sub-channel capacity for THz EM channel.

Because channels are by no way an information source, the upper bound of $C_H$ is $C$ i.e. $C_H \geq C$.

## V. NUMERICAL RESULTS

Here we provide some numerical results on each sub-channel capacity with appropriate simulation parameters and grow some insights into the capacity of the hybrid channel. The capacity of THz EM channel is determined by (3) with the frequency in THz range. The PSD of the transmitted signal is assumed to be flat over the entire bandwidth $B$. Although the noise PSD varies depending upon the transmission distance, we consider flat PSD noise model because of flat PSD signal consideration. Therefore, we have path loss-dependent signal-to-noise ratio (SNR) in (3) and thus the THz sub-channel capacity can be expressed in simpler form as $C_1 = B \times \log_2[1 + SNR \times A(f_i, d_1)]$. Adopting free-space path loss, the sub-channel capacity is presented in Fig. 8. The result suggests that the capacity is highly decreased as the distance approaches on the order of a meter because of high molecular absorption occurrence [9].

Based on (4), we the sub-channel capacity for molecular channel is provided in Fig. 9 using a set of simulation parameters with respective typical values such as system temperature of 300 K, transmit power of 1 pico-watt, distance of 100 micrometers, and diffusion coefficient of $10^{-9}$. The capacity is roughly confined between 2,000 to 3,000 bits/sec with a minimum at around 20 hertz bandwidth.

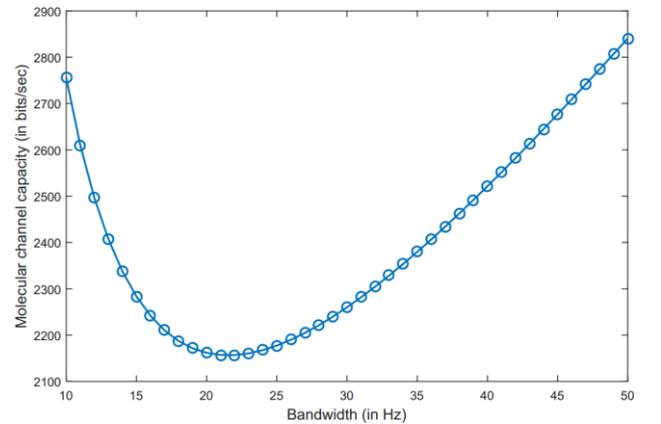

**Fig. 9** Sub-channel capacity for molecular channel.

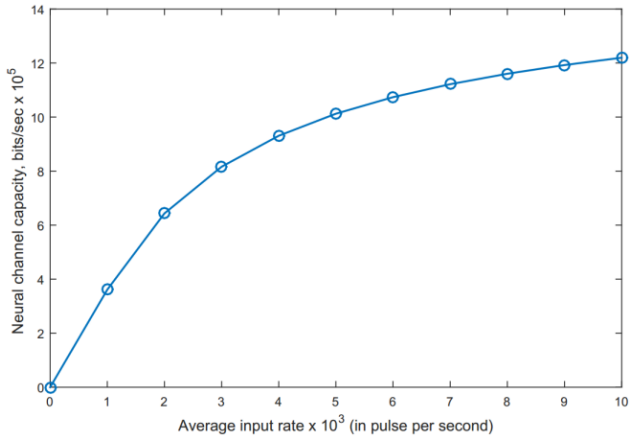

**Fig. 10** Sub-channel capacity for neural channel.

With a neuronal refractory period of 1 millisecond and a latency of 5 microseconds [11], the capacity of a molecular link is represented in Fig. 10. At low input rate the neural sub-channel capacity shows linear trend. However, the capacity approaches to the saturation level as the input increases. We noticed that whereas THz sub-channel and neural sub-channel capacities are comparable, the sub-channel capacity for molecular channel is comparatively too low. The results suggest that the hybrid channel capacity is determined by the molecular channel capacity, according to (7).

## VI. Concluding Remarks

In this paper, we introduce the notion of hybrid nano communication system that is potential to combine three forms of nano communication approaches: terahertz communication molecular communication, and neural communications. It is possible to convert one form of the communications to others by using suitable interfaces. The paper provides some realization on how these interfaces can be designed. Besides, the paper provides a brief analysis of the hybrid channel and each sub-channels. Numerical results show that the capacity of the limit of the hybrid channel is eventually set by the molecular sub-channel capacity.

The paper primarily provides the concept of the HNC system along with its capacity analysis in concise form. The work must come with the analysis of system transfer function in future. In light with that analysis, the capacity analysis should be revisited as well. Various impedance matching associated with the interfaces, neural electrical properties modeling aligned with the HNC concept, and statistical signal processing algorithms for neural signal encoding/decoding are some of the potential research topics that need to be focused on to transform the potentials of the proposed HNC system.